\documentclass[twoside,twocolumn,9pt]{article}
\usepackage{extsizes}
\usepackage[super,sort&compress,comma]{natbib} 
\usepackage[version=3]{mhchem}
\usepackage[left=1.5cm, right=1.5cm, top=1.785cm, bottom=2.0cm]{geometry}
\usepackage{balance}
\usepackage{times,mathptmx}
\usepackage{sectsty}
\usepackage{graphicx} 
\usepackage{lastpage}
\usepackage[format=plain,justification=justified,singlelinecheck=false,font={stretch=1.125,small,sf},labelfont=bf,labelsep=space]{caption}
\usepackage{float}
\usepackage{fancyhdr}
\usepackage{fnpos}
\usepackage[english]{babel}
\usepackage{array}
\usepackage{droidsans}
\usepackage{charter}
\usepackage[T1]{fontenc}
\usepackage[usenames,dvipsnames]{xcolor}
\usepackage{setspace}
\usepackage[compact]{titlesec}
\usepackage{hyperref}

\usepackage{epstopdf}

\definecolor{cream}{RGB}{222,217,201}

\begin{document}

\pagestyle{fancy}
\thispagestyle{plain}
\fancypagestyle{plain}{

}

\makeFNbottom
\makeatletter
\renewcommand\LARGE{\@setfontsize\LARGE{15pt}{17}}
\renewcommand\Large{\@setfontsize\Large{12pt}{14}}
\renewcommand\large{\@setfontsize\large{10pt}{12}}
\renewcommand\footnotesize{\@setfontsize\footnotesize{7pt}{10}}
\makeatother

\renewcommand{\thefootnote}{\fnsymbol{footnote}}
\renewcommand\footnoterule{\vspace*{1pt}%
\color{cream}\hrule width 3.5in height 0.4pt \color{black}\vspace*{5pt}} 
\setcounter{secnumdepth}{5}

\makeatletter 
\renewcommand\@biblabel[1]{#1}            
\renewcommand\@makefntext[1]%
{\noindent\makebox[0pt][r]{\@thefnmark\,}#1}
\makeatother 
\renewcommand{\figurename}{\small{Fig.}~}
\sectionfont{\sffamily\Large}
\subsectionfont{\normalsize}
\subsubsectionfont{\bf}
\setstretch{1.125} 
\setlength{\skip\footins}{0.8cm}
\setlength{\footnotesep}{0.25cm}
\setlength{\jot}{10pt}
\titlespacing*{\section}{0pt}{4pt}{4pt}
\titlespacing*{\subsection}{0pt}{15pt}{1pt}

\fancyfoot{}
\fancyfoot[LO,RE]{\vspace{-7.1pt}\includegraphics[height=9pt]{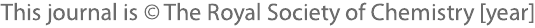}}
\fancyfoot[CO]{\vspace{-7.1pt}\hspace{13.2cm}\includegraphics{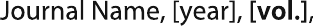}}
\fancyfoot[CE]{\vspace{-7.2pt}\hspace{-14.2cm}\includegraphics{head_foot/RF}}
\fancyfoot[RO]{\footnotesize{\sffamily{1--\pageref{LastPage} ~\textbar  \hspace{2pt}\thepage}}}
\fancyfoot[LE]{\footnotesize{\sffamily{\thepage~\textbar\hspace{3.45cm} 1--\pageref{LastPage}}}}
\fancyhead{}
\renewcommand{\headrulewidth}{0pt} 
\renewcommand{\footrulewidth}{0pt}
\setlength{\arrayrulewidth}{1pt}
\setlength{\columnsep}{6.5mm}
\setlength\bibsep{1pt}

\makeatletter 
\newlength{\figrulesep} 
\setlength{\figrulesep}{0.5\textfloatsep} 

\newcommand{\topfigrule}{\vspace*{-1pt}%
\noindent{\color{cream}\rule[-\figrulesep]{\columnwidth}{1.5pt}} }

\newcommand{\botfigrule}{\vspace*{-2pt}%
\noindent{\color{cream}\rule[\figrulesep]{\columnwidth}{1.5pt}} }

\newcommand{\dblfigrule}{\vspace*{-1pt}%
\noindent{\color{cream}\rule[-\figrulesep]{\textwidth}{1.5pt}} }

\makeatother

\twocolumn[
  \begin{@twocolumnfalse}
\vspace{3cm}
\sffamily
\begin{tabular}{m{4.5cm} p{13.5cm} }

\includegraphics{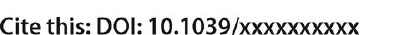} & \noindent\LARGE{\textbf{Complexes of gold and imidazole formed in helium nanodroplets}} \\
\vspace{0.3cm} & \vspace{0.3cm} \\

 & \noindent\large{Michael Gatchell,$^{\ast}$\textit{$^{ab}$} Marcelo Goulart,\textit{$^{a}$} Lorentz Kranabetter,\textit{$^{a}$} Martin Kuhn,\textit{$^{a}$} Paul Martini,\textit{$^{a}$} Bilal Rasul,\textit{$^{ac}$} and Paul Scheier\textit{$^{a}$}} \\

\includegraphics{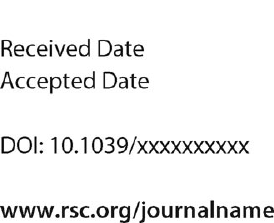} & \noindent\normalsize{We have studied complexes of gold atoms and imidazole (C$_3$N$_2$H$_4$, abbreviated Im) produced in helium nanodroplets. Following the ionization of the doped droplets we detect a broad range of different Au$_m$Im$_n^+$ complexes, however we find that for specific values of $m$ certain $n$ are ``magic'' and thus particularly abundant. Our density functional theory calculations indicate that these abundant clusters sizes are partially the result of particularly stable complexes, e.g.\ AuIm$_2^+$, and partially due to a transition in fragmentation patterns from the loss of neutral imidazole molecules for large systems to the loss of neutral gold atoms for smaller systems.} \\

\end{tabular}

 \end{@twocolumnfalse} \vspace{0.6cm}

  ]

\renewcommand*\rmdefault{bch}\normalfont\upshape
\rmfamily
\section*{}
\vspace{-1cm}


\footnotetext{\textit{$^{a}$~Institut f\"{u}r Ionenphysik und Angewandte Physik, Universit\"{a}t Innsbruck, Technikerstr.~25, A-6020 Innsbruck, Austria; E-mail: michael.gatchell@uibk.ac.at}}
\footnotetext{\textit{$^{b}$~Department of Physics, Stockholm University, 106 91 Stockholm, Sweden; E-mail: gatchell@fysik.su.se}}
\footnotetext{\textit{$^{c}$~Department of Physics, University of Sargodha, 40100 Sargodha, Pakistan}}





\section{Introduction}
Gold at the molecular level is a completely different beast than the bulk material. The metal is known for its chemical stability which gives rise to its untarnished luster, but atomic gold has a complex chemistry that goes far beyond metal alloys. Gold's chemical properties are a result of its 5d$^{10}$6s$^1$ valence electronic structure which, due to gold being a third row transition metal, is strongly influenced by relativistic effects \cite{Schwarz:2003ip}. This leads to a contraction of the 6s orbital which gives the gold atom both a high ionization potential ($\text{IE}=9.25$\,eV \cite{Brown:1978aa}) and electron affinity ($\text{EA}= 2.31$\,eV \cite{Andersen:1999aa}). The stabilization of the 6s orbital also leads to gold atoms exhibiting chemical properties similar to ``big'' hydrogen atoms in a number of chemical environments.\cite{Zeller:1991aa,Schmidbaur:1992aa,Angermaier:1994aa,Schmidbaur:1995aa,Benitez:2009aa,Wang:2010aa} 

Clusters and nanoparticles of gold in particular have been shown to be remarkably good catalysts, debunking the previous assumption that gold had little chemical activity.\cite{Hashmi:2006fr,Hashmi:2007hl,Corma:2008fy} Much of this focus has been on the roll of gold complexes in driving hydrogenation and oxidization reactions in organic chemistry.\cite{Hashmi:2006fr,Hashmi:2007hl,Corma:2008fy} A category of such catalysts that have found interest recently are gold nanoclusters protected by organic ligands, such as Au$_{m}$(SPh)$_{n}$ (Ph = C$_6$H$_5$) systems, where a core consisting of a Au$_8^{4+}$ cluster is stabilized by the surrounding ligands.\cite{Li:2014ga,Takagi:2015hx} These types of complexes are used as catalysts for effective site-selective hydrogenation.\cite{Li:2014ga,Takagi:2015hx}


Organometallic complexes first sparked significant interest with the discovery of the ferrocene in the 1950s.\cite{Kealy:1951aa,Miller:1952aa,Werner:2012aa} Work in this field has since continued with the interactions between metals and organic systems being deduced from studies of their interactions with individual molecular building blocks. However, relatively little work has been done investigating the interactions between gold---in particular gold clusters---and many groups of organic molecules. New results on this matter can help improve our understanding as to how gold interacts with biological systems and may also play a role in the development of new gold-based catalysts. This work is the first part of a series of studies we are carrying out to investigate the properties of gold nanoparticles in various chemical environments.




\begin{figure}[h]
\begin{center}
\includegraphics[width=3cm]{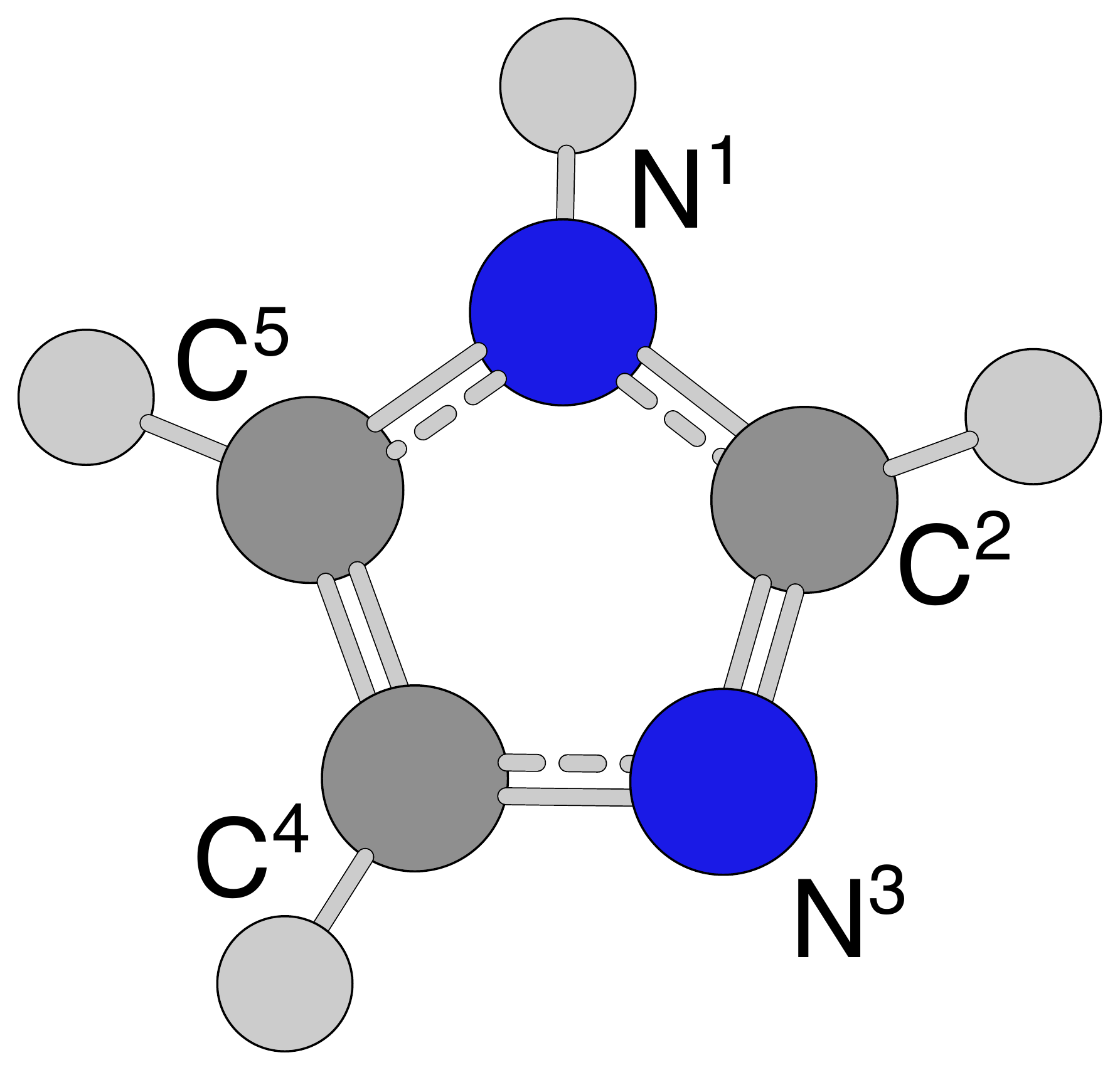}
\caption{The molecular structure of imidazole. The labels next to each C and N atom are used to denote the locations of bond sites when discussing our results.\label{fig:Im0}}
\label{default}
\end{center}
\end{figure}

\begin{figure*}[h]
\begin{center}
\includegraphics[width=1\textwidth]{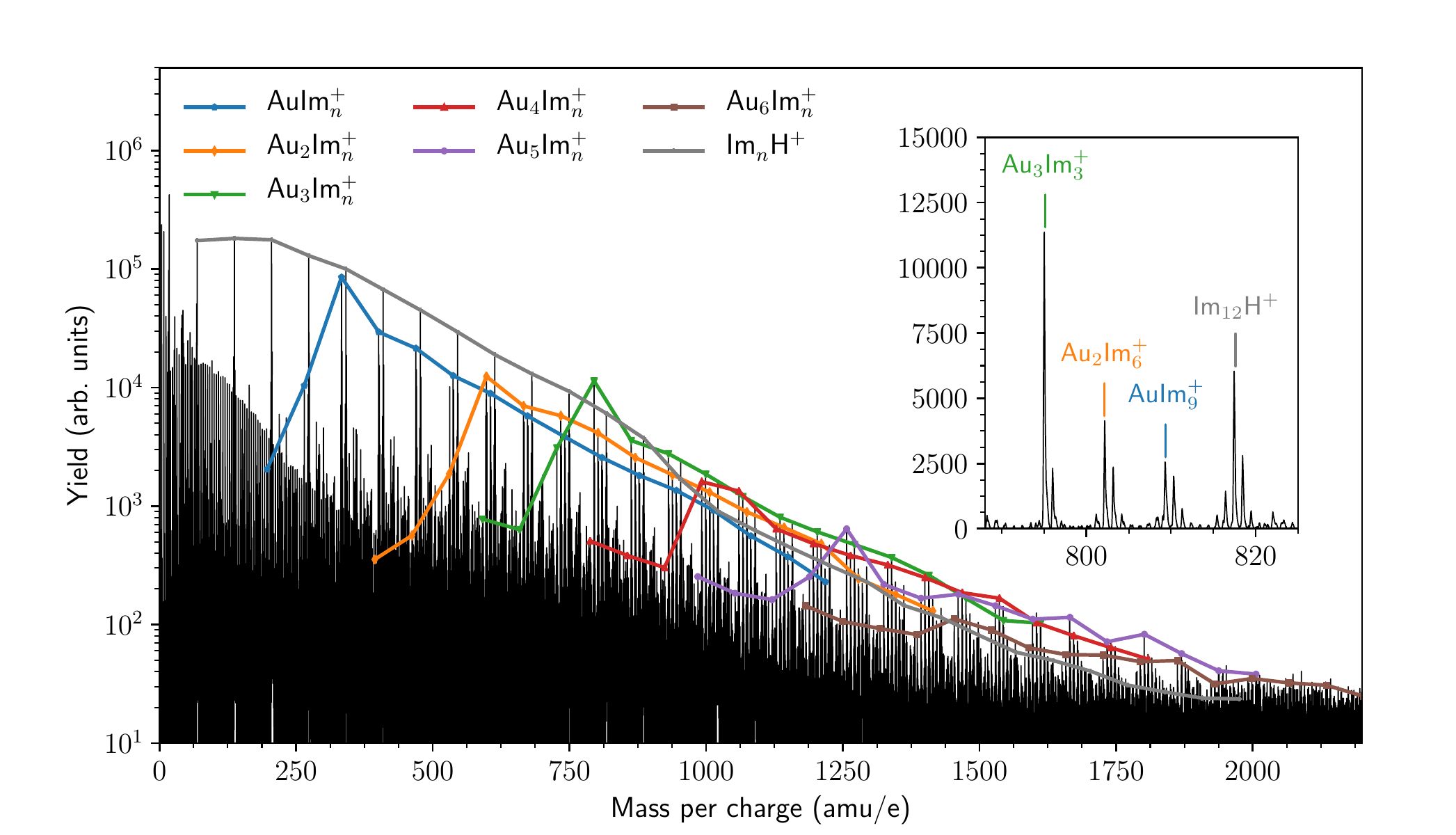}
\caption{Cationic mass spectrum of ionized helium nanodroplets containing gold and imidazole. Series of Au$_m$Im$_n^+$ with fixed values of $m$ and $0\leq n \leq 15$ are indicated by their peak heights and grouped by color. The pure imidazole clusters (gray line) are predominantly protonated and follow a smooth size distribution. The extra proton likely originates from the dissociation of an imidazole unit in a larger neutral precursor cluster upon charge transfer from He$^+$. Clusters containing gold are not observed to be protonated. The inset shows a zoom in of the mass spectrum where four peaks (the main peak in each isotopologue series) are labeled, demonstrating the high resolution of the current setup.\label{fig:AuImMSraw}}
\label{default}
\end{center}
\end{figure*}

In this work we have studied complexes consisting of one or more gold atoms/ions and imidazole (C$_3$N$_2$H$_4$, abbreviated Im) molecules. Imidazole is a simple nitrogen-containing heterocycle consisting of the pentagonal structure shown in Figure \ref{fig:Im0}. An aromatic molecule, imidazole forms the basis for a wide range of organic molecules such as the DNA bases guanine and adenine, histidines, histamines, and the aforementioned NHC complexes to name a few.\cite{Kuhn:2018aa} We produce neutral Au$_m$Im$_n$ complexes in superfluid helium nanodroplets which are then ionized. Using mass spectrometry to study the Au$_m$Im$_n^+$ products, we identify a number of seemingly ``magic'' combinations of $m$ and $n$, i.e.\ structures that are particularly abundant compared to those with $n\pm1$ imidazole molecules for a given $m$ Au atoms. We have also performed quantum chemical structure calculations to identify the origin of these abundant structures and find that certain structures, e.g.\ AuIm$_2^+$ and Au$_3$Im$_3^+$, are indeed much more stable than systems with more imidazole molecules. We also find that clusters larger (i.e.\ containing more imidazole rings) than the most abundant sizes predominantly dissociate through the loss of neutral imidazole molecules, while at and below the these sizes (i.e.\ fewer imidazoles) they mainly lose one or more neutral gold atoms.

\section{Experimental methods}
The experiments carried out for this work were performed using the apparatus described in detail in Refs.\cite{Schobel:2011aa,Kurzthaler:2016aa,Kuhn:2016aa} Here, superfluid helium nanodroplets are formed in the supersonic expansion of He gas at a pressure of 2.5\,Mpa into a vacuum through the 5\,$\mu$m diameter aperture of a nozzle cooled to 9.4\,K. This results in a broad size distribution of He droplet sizes with a mean size of approximately 10$^6$ atoms. A beam of He nanodroplets is formed and the center part passes a skimmer with a 0.8\,mm diameter aperture positioned 8\,mm from the nozzle. The beam then traverses two sequential pickup chambers where the droplets are doped with gaseous imidazole molecules and Au atoms, respectively. The imidazole vapor is formed in an oven containing imidazole powder ($>99$\% purity from Sigma Aldrich) operating at 420\,K, and in the second chamber an oven heated to approximately 1300\,K is used to vaporize solid gold. The number of imidazole molecules and Au atoms picked up by each droplet varies following a Poisson distribution, with the average number of dopants in each individual droplet depending on its geometrical cross section. In the next stage the doped droplets are ionized by irradiation with electrons with a kinetic energy of 88\,eV. Positively charged products are analyzed using a reflectron time-of-flight mass spectrometer from Tofwerk AG (model HTOF). The raw mass spectrum is reduced using the IsotopeFit software,\cite{Ralser:2015aa} which deconvolutes overlapping features lying close to each other in mass (such as different cluster makeups with the same nominal mass).

\section{Theoretical methods}

We have studied the structures and stabilities of gold and imidazole complexes using Self Consistent Charge Density Functional Tight Binding (SCC-DFTB) and Density Functional Theory (DFT) calculations. Due to the high speed at which SCC-DFTB calculations can be performed, we used this method to screen input structures before running more expensive DFT structure optimizations. For this we used the auorg-1-1 parameter set.\cite{Fihey:2015aa,PhysRevB.58.7260} For the DFT calculations we used the M06 functional---which has been shown to give good results for organic complexes containing transition metals \cite{Zhao:2009aa}---together with the Zeroth Order Regular Approximation (ZORA) relativistic correction.\cite{Lenthe:1993aa,Lenthe:1999aa} Structures were optimized with the M06 functional and an all-electron def2-SVP basis set before single point energy calculations were performed with a def2-TZVP basis set.


The SCC-DFTB calculations were performed using the DFTB+ package (release 17.1) \cite{Aradi:2007aa} and the DFT calculations were run using the ORCA suite (version 4.0.1).\cite{Neese:2012aa}

\section{Results}
A mass spectrum of cationic species formed when helium droplets containing gold and imidazole are ionized is shown in Figure \ref{fig:AuImMSraw}. Here we have highlighted the highest peaks from the Au$_m$Im$_n^+$ ($1\leq m \leq6$ and $0\leq n \leq 15$) cluster series. All of these complexes are bare of any helium, indicating that these ions have been pushed out of a droplet by other charged species that are formed at the same time. A consequence of the exothermic charge transfer from He$^+$, formed when the droplet is ionized, to the dopant is that the neutral gold and imidazole complexes that are initially in the droplets are expected to undergo restructuring---such as partial dissociation. 

\begin{figure}[h]
\begin{center}
\includegraphics[width=1\columnwidth]{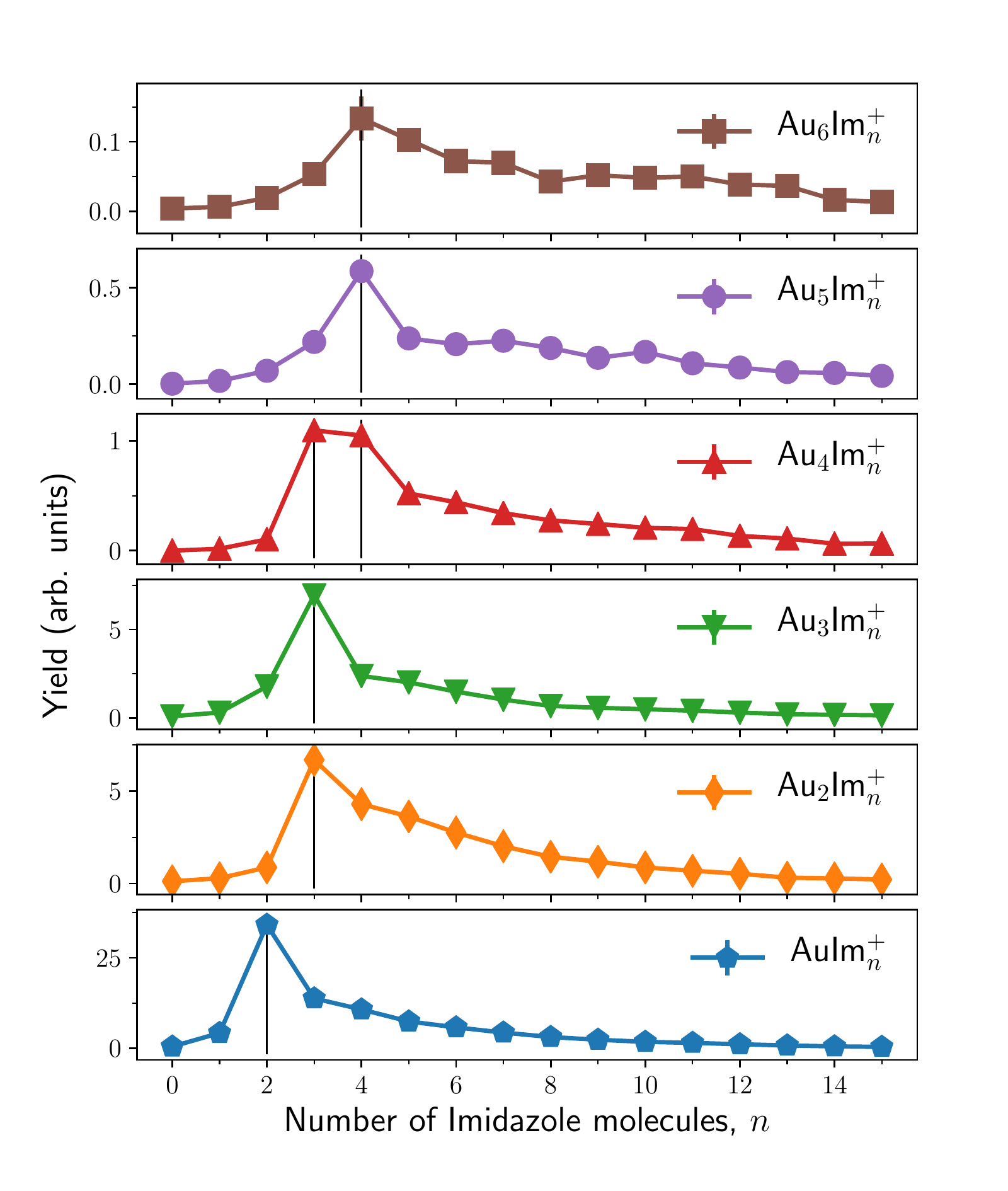}
\caption{Integrated intensities of Au$_m$Im$_n^+$ complexes arranged by fixed values of $m$. Vertical lines indicate the most abundant number of imidazole molecules in each panel. The statistical error bars are in most cases smaller than the data points. \label{fig:AuImMS}}
\label{default}
\end{center}
\end{figure}

Figure \ref{fig:AuImMS} shows an overview of the integrated intensities of Au$_m$Im$_n^+$ complexes from the mass spectrum in Figure \ref{fig:AuImMSraw}. In each of the panels of Figure \ref{fig:AuImMS} (each number of Au atoms) we see that a specific number of imidazole molecules ($n=2$ in the case of AuIm$_n^+$) is significantly more abundant than the rest for a given gold cluster size. Complexes with a single Au atom seem to prefer to contain two imidazole molecules while those with 2 or 3 Au atoms preferably carry 3 imidazole molecules. Systems with 4 Au atoms stand out with two strong features coming from complexes with 3 or 4 imidazole molecules. Finally, the largest systems studied here, with 5 and 6 Au atoms, mainly contain 4 imidazole rings. Other combinations of gold and imidazole have significantly lower abundances, with broad distributions that could be remnants of the size distributions prior to ionization. There is also a trend that odd-numbered gold clusters ($n=1,3,5$) show somewhat more prominent intensity maximas that the even-numbered clusters ($n=2,4,6$) which could indicate that the standout odd-numbered structures are particularly stable. Due to the processes in which the cations are formed and the behavior of the mass spectra, we expect the ``magic'' sizes in each series to mainly be produced through the decay of larger counterparts (e.g.\ those with more imidazole molecules) instead of forming in the neutral droplets. It is thus difficult to directly compare the intensities of complex sizes above and below the maxima in each cluster series as this appears to be a bottle-neck in the decay process.

We have investigated the structures of AuIm$_n^+$, where $n=1,2,3$; Au$_2$Im$_n^+$, where $n=1,2,3,4$; Au$_3$Im$_n^+$, where $n=1,2,3,4$; and Au$_4$Im$_n^+$, where $n=1,2,3,4,5$, using DFT structure calculations in order to better understand the experimental results. We have performed calculations on several geometries for each cluster size, and the geometries that are shown here are those with the lowest energies and those that lie within 0.15\,eV of the lowest energy structure (other than a few exceptions for education purpose, e.g.\ the different AuIm$_m$ geometries). The dissociation energy given with each structure in Figures \ref{fig:Au1Imn} through \ref{fig:Au4Imn} is the lowest dissociation energy found for that cluster geometry. The corresponding dissociation pathways are discussed in the text and summarized in Figure \ref{fig:AuIm_matrix}. The results of the calculations are covered in the next sections and a summary is given in the discussion.

\subsection{AuIm$_n^+$}

\begin{figure}[h]
\begin{center}
\includegraphics[width=\columnwidth]{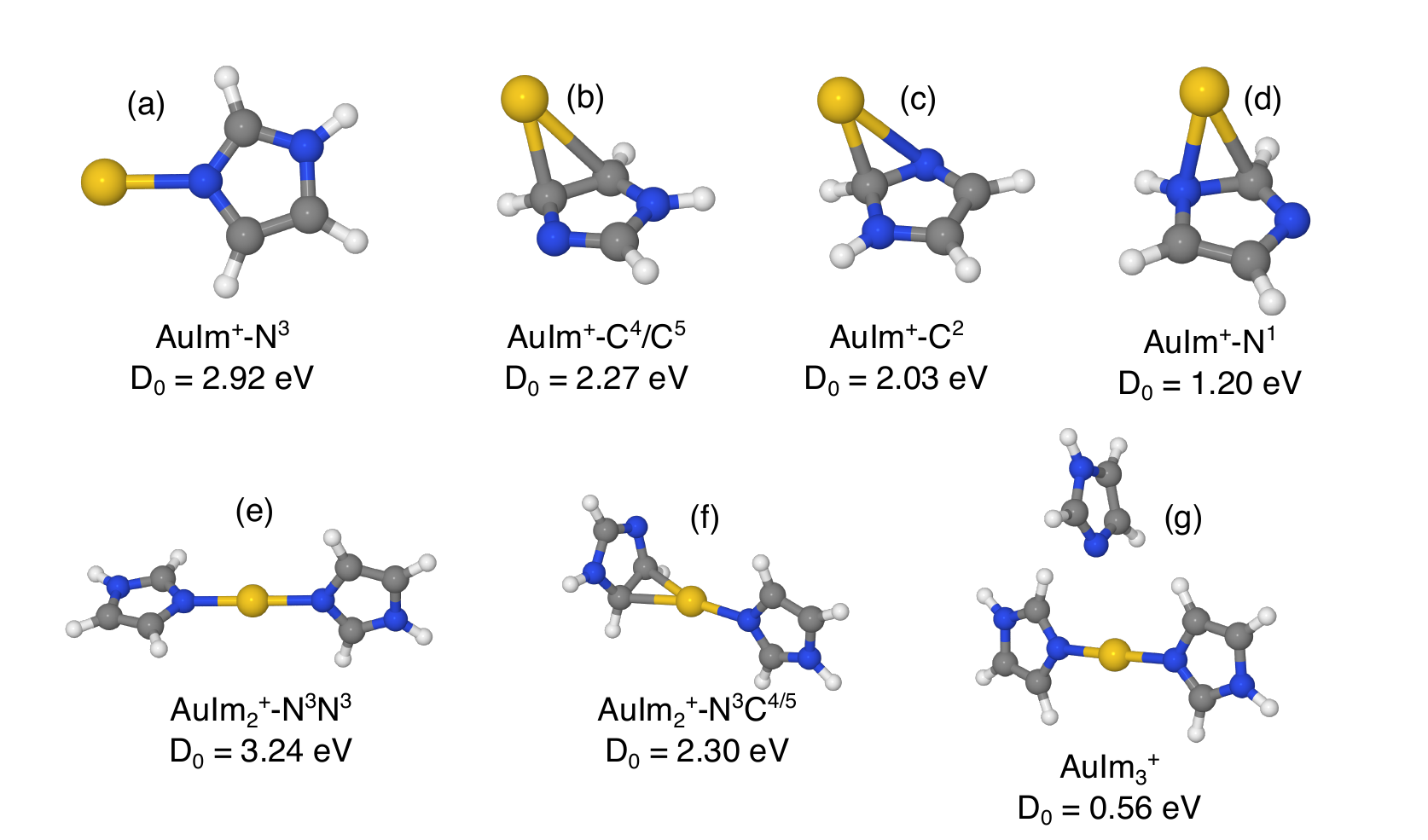}
\caption{Proposed structures of AuIm$_{n}^+$, $n=1,2,3$, and lowest dissociation energies from M06/def2-TZVP//M06/def2-SVP calculations.  \label{fig:Au1Imn}}
\label{default}
\end{center}
\end{figure}

The lowest energy structure of AuIm$^+$ is found when Au forms a covalent bond to the N$^3$ site of the imidazole ring, forming the planar lollipop-shaped structure labelled {\bf (a)} in Figure \ref{fig:Au1Imn}. The calculated binding energy of this system is 2.9\,eV for the AuIm$^+ \rightarrow$ Im$^+$ + Au dissociation channel. In addition to this structure, three other geometries identified for the AuIm$^+$ are shown as panels {\bf (b)}--{\bf (d)}. In each of the latter cases, the Au is located above the plane and near the edge of the imidazole ring with a binding energy that is significantly lower than bond at the N$^3$ position. 

This trend is repeated with the addition of a second imidazole molecule (structures {\bf (e)} and {\bf (f)} in Figure \ref{fig:Au1Imn}). The second ring preferentially forms a bond via the N$^3$ atom to the Au center, resulting in a dumbbell shaped ImAuIm$^+$ structure (panel {\bf (e)}). Alternative structures, like the one shown in panel {\bf (f)}, follow the same order in binding energies as in the case of AuIm$^+$. Noteworthy is that the binding energy of the second imidazole ring is \emph{higher} than that of the first one by a few tenths of an eV for the most stable structures. 

The lowest energy AuIm$_3^+$ system found in our calculations is shown in panel {\bf (g)} in Figure \ref{fig:Au1Imn}. Compared to the first two rings, the third imidazole is only weakly bound to the Au core, in agreement with the experimental results where AuIm$_{n>2}^+$ are much less abundant than AuIm$_{2}^+$ complexes. 


\subsection{Au$_2$Im$_n^+$}

\begin{figure}[h]
\begin{center}
\includegraphics[width=\columnwidth]{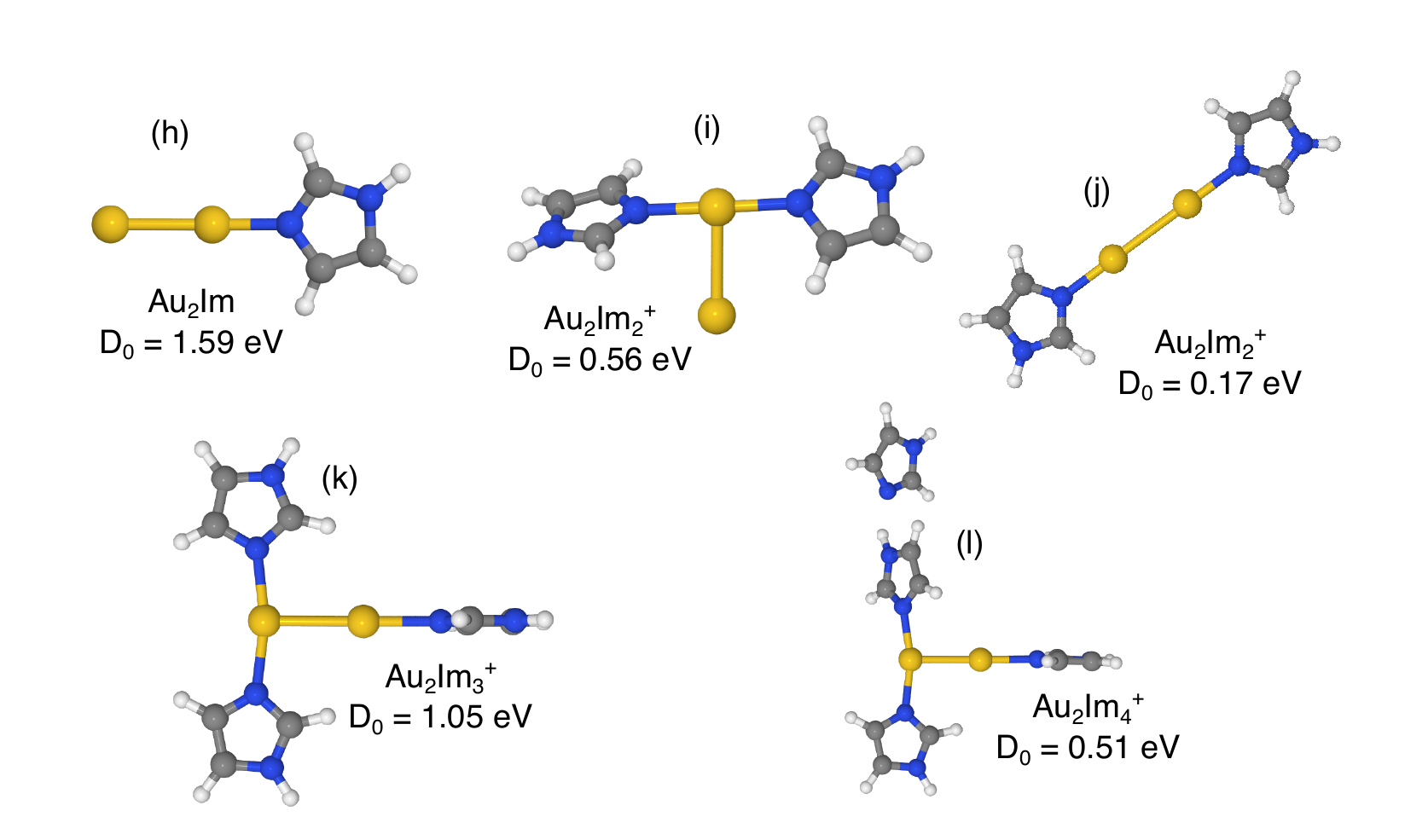}
\caption{Proposed structures of Au$_2$Im$_{n}^+$, $n=1,2,3,4$, and lowest dissociation energies from M06/def2-TZVP//M06/def2-SVP calculations.  \label{fig:Au2Imn}}
\label{default}
\end{center}
\end{figure}

In Figure \ref{fig:Au2Imn} we show optimized structures for Au$_2$Im$_n^+$ complexes where $n=1,2,3,4$. For Au$_2$Im$^+$ and Au$_2$Im$_2^+$ (structures {\bf (h)} and {\bf (i)}) the trends observed for systems with a single Au are repeated. In both of these cases the lowest energy dissociation channel is the loss of a neutral Au atom, resulting in structures shown in Figure \ref{fig:Au1Imn}. The imidazole molecules thus contribute to breaking down the gold clusters. Structure {\bf (j)} in Figure \ref{fig:Au2Imn} shows an alternative form of Au$_2$Im$_2^+$ with a dumbbell shape which is significantly less stable than structure {\bf (i)}, which is mainly due to a weakening of the Au-Au bond. 

The addition of a third imidazole ring (structure {\bf (k)}) stabilizes the Au$_2$Im$_3^+$ system relative to clusters with only two rings. The lowest energy structure that we have found forms a T-shape, with two rings attached to one Au atom and a single ring to the other. There are two competing dissociation pathways for this system separated by less than 0.1\,eV, the Au$_2$Im$_3^+ \rightarrow$ Au$_2$Im$_2^+$ + Im and Au$_2$Im$_3^+ \rightarrow$ AuIm$_2^+$ + AuIm, with the prior having the lower dissociation energy. 

We have identified a few different Au$_2$Im$_4^+$ structures, all with the same properties and similar binding energies. The lowest energy isomer found is shown as structure {\bf (l)}. Here the fourth imidazole ring forms a weak bond at a N-H-N site, essentially a hydrogen bond, with one of the other rings. Other positions are possible too, however none results in the fourth imidazole ring forming a covalent bond with the Au$_2$ core.

\subsection{Au$_3$Im$_n^+$}

\begin{figure}[h]
\begin{center}
\includegraphics[width=\columnwidth]{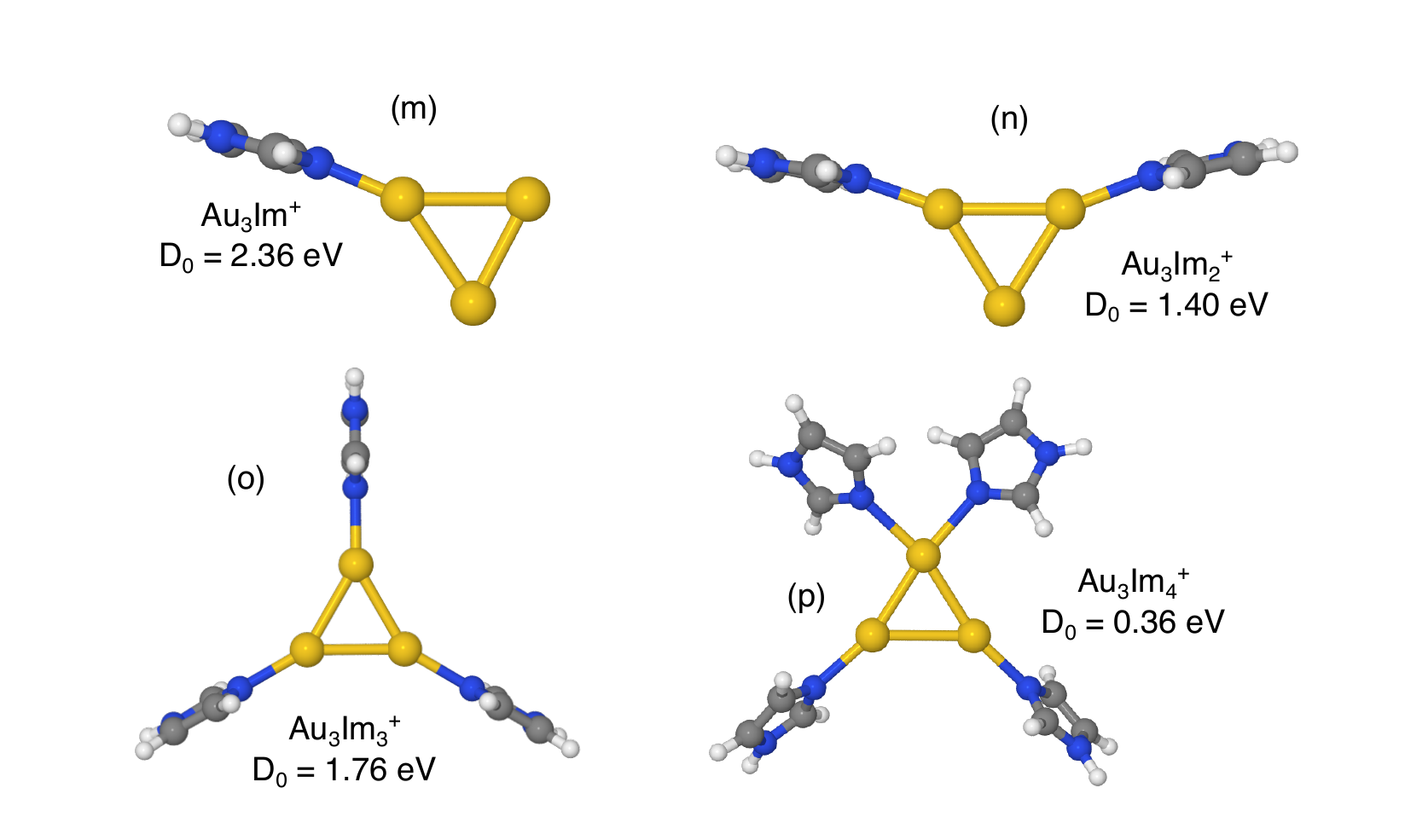}
\caption{Proposed structures of Au$_3$Im$_{n}^+$, $n=1,2,3,4$, and lowest dissociation energies from M06/def2-TZVP//M06/def2-SVP calculations.  \label{fig:Au3Imn}}
\label{default}
\end{center}
\end{figure}

The cationic gold trimer forms an equilateral triangle shaped structure \cite{Schooss:2010aa} and this basic structure is only weakly influenced by the binding of an imidazole ring (structure {\bf (m)} in Figure \ref{fig:Au3Imn}). Again the imidazole binds to the metal cluster via the N$^3$ site and the ring orients itself perpendicularly to the plane of the cluster. The lowest energy dissociation pathway here is the loss of the neutral imidazole ring leaving the bare Au$_3^+$ cluster. Similarly, the lowest energy structure for Au$_3$Im$_2^+$ that we have identified, structure {\bf (n)}, has a triangular Au$_3$ core with the two imidazole rings occupying separate Au atoms. The dissociation energy for Au$_3$Im$_2^+$ (about 1.4 eV) is somewhat lower than for Au$_3$Im$^+$ and Au$_3$Im$_3^+$ (2.4 and 1.8 eV, respectively for structures {\bf (m)} and {\bf (o)}), with the Au$_3$Im$_2^+ \rightarrow$ AuIm$_2^+$ + Au$_2$ pathway giving the lowest dissociation energy.

Like with the Au$_2$Im$_3^+$ systems, Au$_3$Im$_3^+$ (structure {\bf (o)}) is stabilized by the presence of a third imidazole molecule resulting in as structure with near C$_3$ symmetry (depending on the orientation of the imidazole rings). The lowest energy dissociation channel for Au$_3$Im$_3^+$ is through the loss of a neutral imidazole ring resulting in Au$_3$Im$_2^+$ as the charged fragment.

A proposed structure of Au$_3$Im$_4^+$ is shown as panel {\bf (p)} in Figure \ref{fig:Au3Imn}. The fourth imidazole interacts here directly with the Au$_3$ core, however it lowers the overall stability of the cluster such that the dissociation energy for losing a neutral imidazole ring is only about 0.4 eV.

\subsection{Au$_4$Im$_n^+$}
\begin{figure}[h]
\begin{center}
\includegraphics[width=\columnwidth]{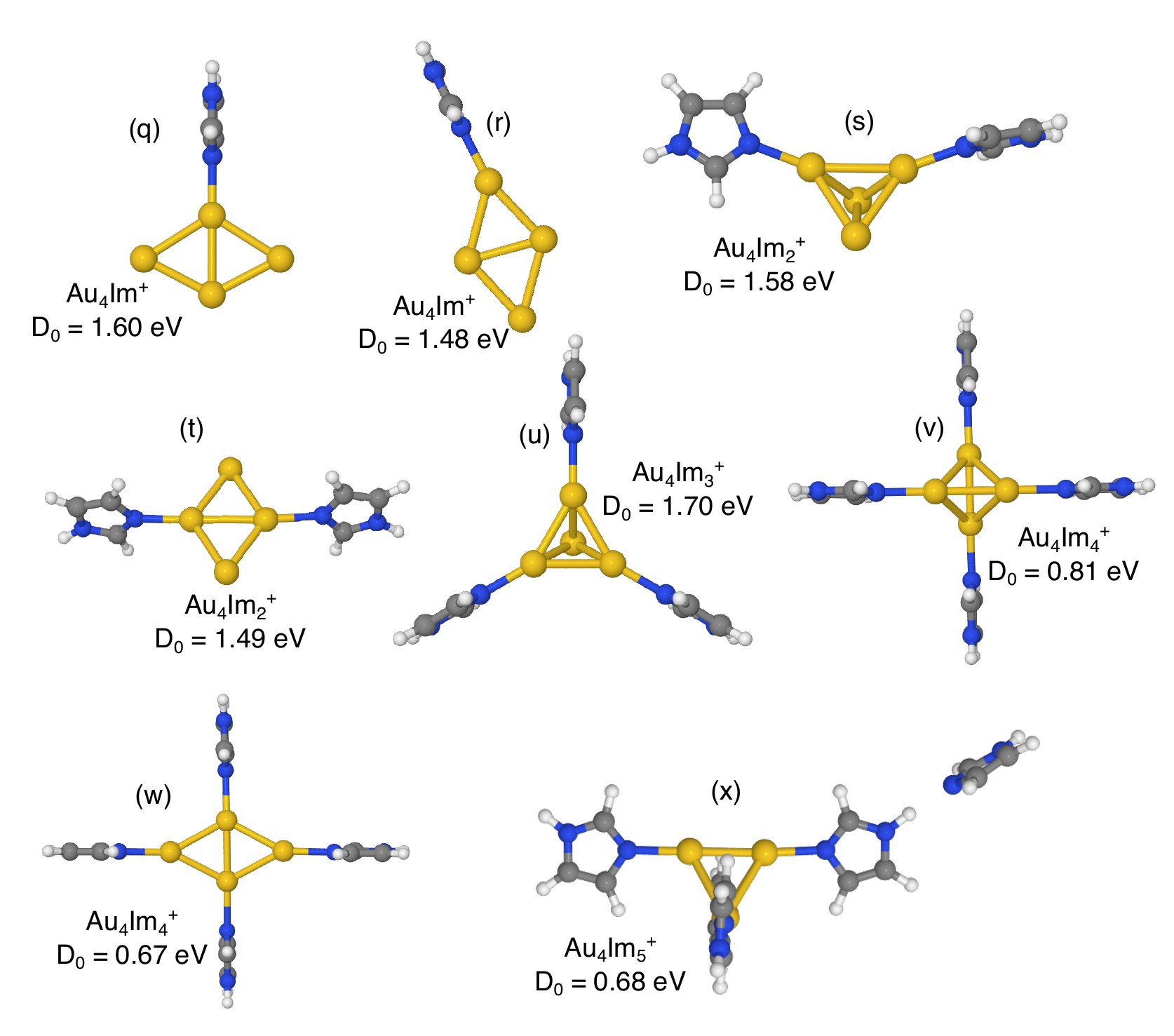}
\caption{Proposed structures of Au$_4$Im$_{n}^+$, $n=1,2,3,4,5$, and lowest dissociation energies from M06/def2-TZVP//M06/def2-SVP calculations. \label{fig:Au4Imn}}
\label{default}
\end{center}
\end{figure}

The most stable isomer of Au$_4^+$ is a rhombic structure \cite{Schooss:2010aa}, with a tetrahedron structure lying about 0.3\,eV above the prior according to our calculations. The rhombic structure remains energetically preferred following the addition of an imidazole molecule, with two possible structures ({\bf (q)} and {\bf (r)} in Figure \ref{fig:Au4Imn}) being separated by about 0.1\,eV. The roles are reversed when two imidazole rings bind to Au$_4^+$. Now the decorated tetrahedron structure is preferred over the rhombic structure. The lowest energy examples of these two types of structures are labelled {\bf (s)} and {\bf (t)} in Figure \ref{fig:Au4Imn}. Additional complexes with a rhombic Au$_4$ substructures and different locations of the two imidazole rings have potential energies more than 0.1\,eV higher than structure {\bf (t)}.

Our calculations indicate that the Au$_4$Im$_3^+$ structure strongly prefers the tetrahedron geometry ({\bf (u)} in Figure \ref{fig:Au4Imn}). Attempts to optimize Au$_4$Im$_3^+$ with a rhombic Au$_4$ substructure all result in the folding of the gold atoms into a tetrahedron. The most stable Au$_4$Im$_3^+$ structure has a threefold symmetry similar to Au$_3$Im$_3^+$ and the lowest energy dissociation pathway identified is the loss of a single neutral Au atom. For Au$_4$Im$_4^+$ we identify both tetrahedron {\bf (v)} and rhombic {\bf (w)} geometries, with the prior being energetically preferred. The tetrahedron Au$_4$Im$_4^+$ consists of two linear Au$_2$Im$_2$ elements arranged perpendicularly to each other with their axes offset by about 2.25\,\AA. The most stable Au$_4$Im$_5^+$ complex we have identified is labeled {\bf (x)} in Figure \ref{fig:Au4Imn}. This consists of the basic tetrahedron Au$_4$ core decorated with four imdazoles, structure {\bf (v)}, with the fifth ring forming a hydrogen bond with one of the NH sites of a neighboring ring. It is the loss of the fifth ring which has the lowest dissociation energy for this system. Similar structures are observed with a rhombic Au$_4$ core, although these are less stable by at least a few tenths of an eV.

\subsection{Au$_5$Im$_n^+$ and Au$_6$Im$_n^+$}

Due to the increasing complexity and number of isomers, we have not performed calculations for systems with five or more Au atoms. 

\section{Discussion and Conclusions}

\begin{figure}[h]
\begin{center}
\includegraphics[width=\columnwidth]{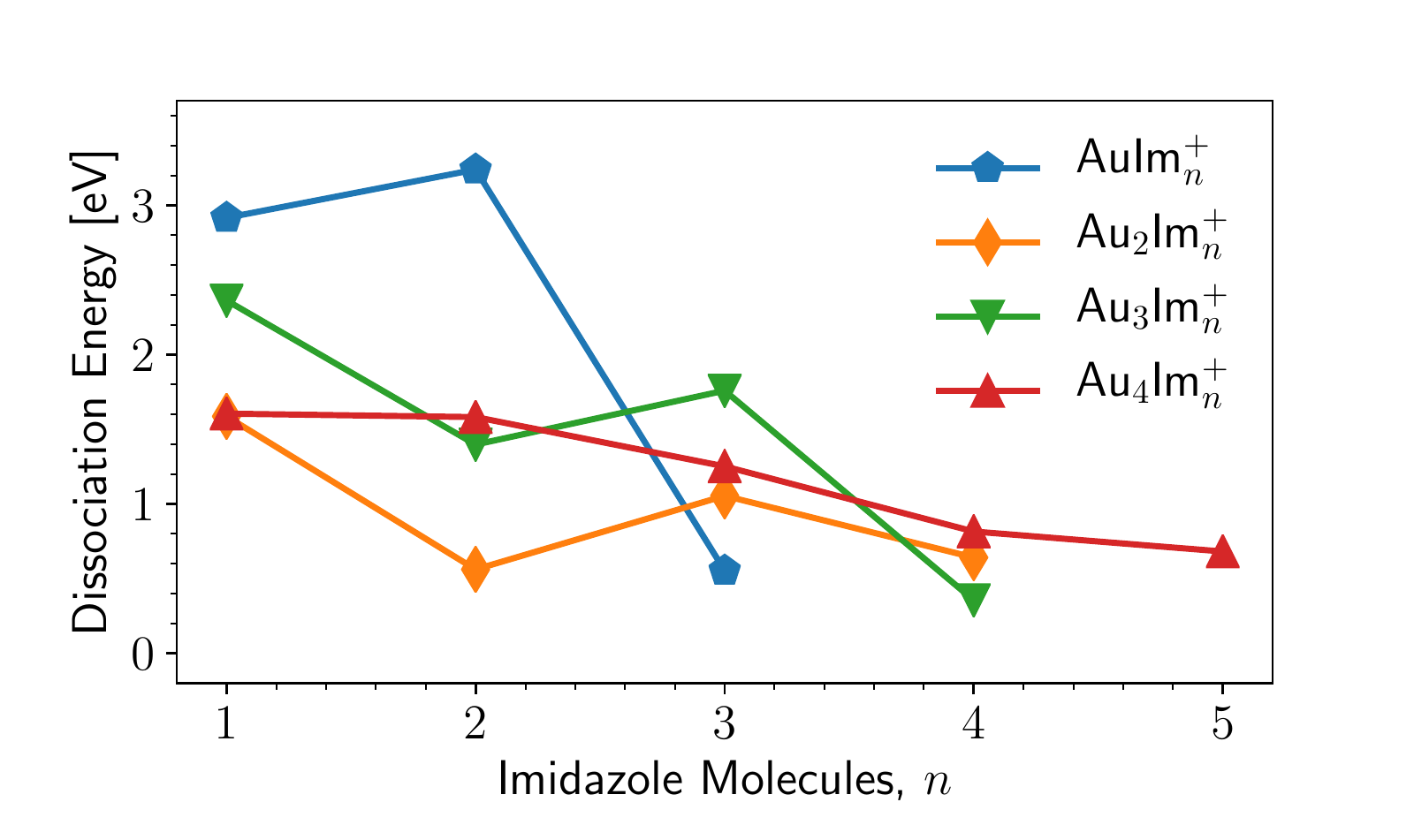}
\caption{Lowest dissociation energies determined by our structure calculations at M06/def2-TZVP//M06/def2-SVP level of theory. The lines connecting the data points are to guide the reader. The corresponding charged fragments are given in Figure \ref{fig:AuIm_matrix}.  \label{fig:AuIm_DFT_BEs}}
\label{default}
\end{center}
\end{figure}

A summary of the calculated dissociation energies of Au$_m$Im$_n^+$ complexes is shown in Figure \ref{fig:AuIm_DFT_BEs}. From this figure it is clear that one of the ``magic'' structures observed in the experiments (Figure \ref{fig:AuImMS}), AuIm$_2^+$ and Au$_3$Im$_3^+$, stand out as being significantly more stable than clusters containing one additional imidazole ring. However, the Au$_2$Im$_3^+$, Au$_4$Im$_3^+$,and Au$_4$Im$_4^+$ systems are only slightly more stable than those containing one additional imidazole ring. It is thus not clear from the dissociation energies alone why these complexes are so abundant in our experiments relative to other systems of similar size.

\begin{figure}[h]
\begin{center}
\includegraphics[width=0.75\columnwidth]{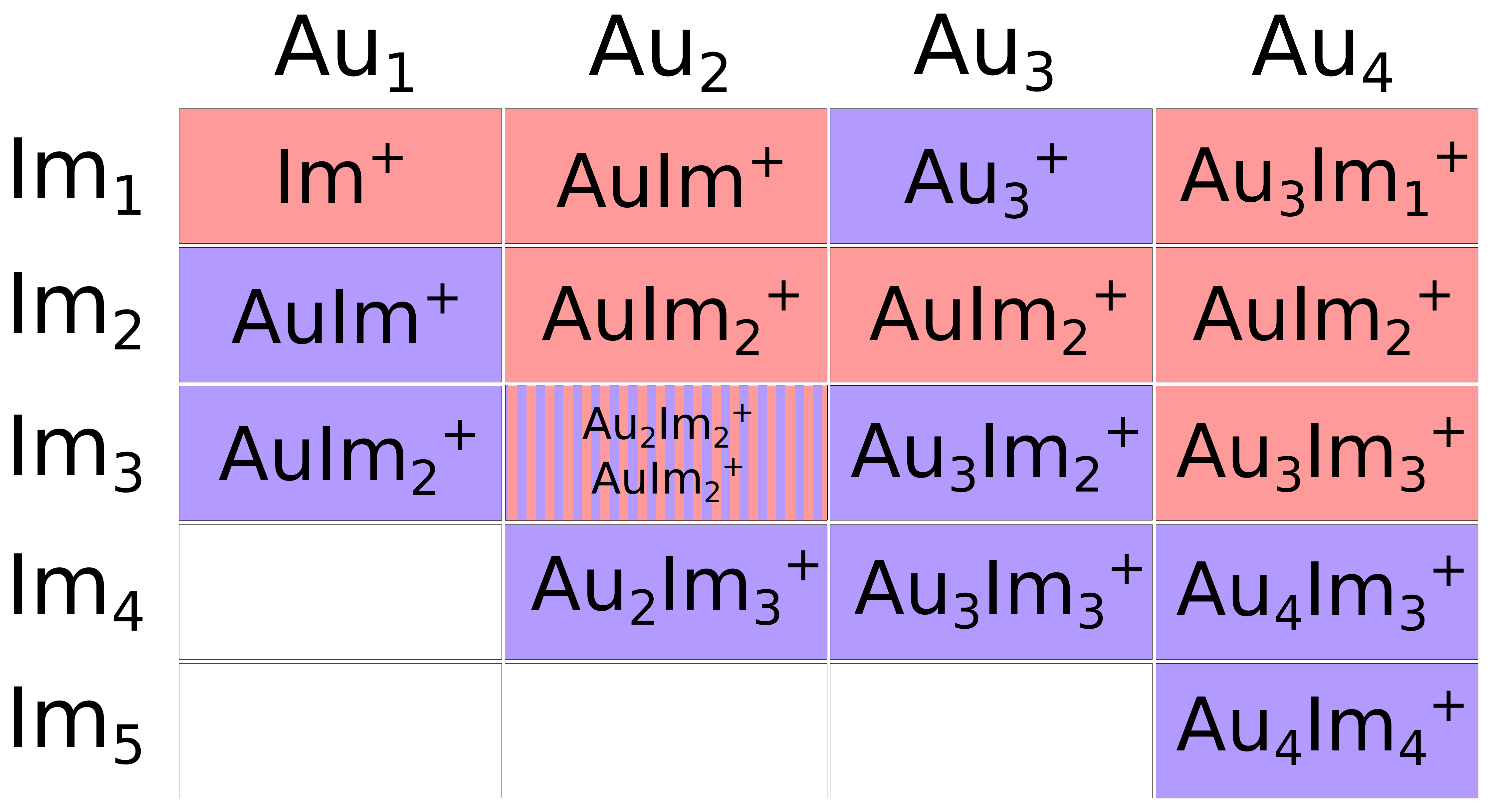}
\caption{Matrix indicating the charged product that remains when Au$_m$Im$_n^+$ dissociates along the lowest energy pathway (not considering intermediate barriers). Columns represent fixed values of $m$ and rows values of $n$. Cells shaded red indicate products that have loss at least one Au atom (and any number of imidazole), while blue cells indicate the loss of a single neutral imidazole ring. The striped cell for the Au$_2$Im$_3^+$ parent indicates two dissociation energies separated by less than 0.1\,eV. We expect all systems larger than those shown here to fragment mainly through the loss of neutral imidazole. \label{fig:AuIm_matrix}}
\label{default}
\end{center}
\end{figure}

A contributing factor to the observed ``magic'' numbers in the experiments is likely the specific fragmentation patterns exhibited when the different systems dissociate. In Figure \ref{fig:AuIm_matrix} we present the charged products produced when each Au$_m$Im$_n^+$ cluster size dissociates along its lowest energy pathway (disregarding potential reaction barriers). For large systems, with more imidazole rings than the observed maxima in each gold cluster series, the lowest energy dissociation pathway is consistently the loss of neutral, loosely bound imidazole molecules. After losing a sufficient number of imidazole molecules, systems that remain hot enough can then dissociate through the loss of neutral gold atoms, possibly together with one or more imidazole rings. This transition between losing neutral imidazole and neutral gold correlates well with the experimentally observed ``magic'' number of imidazole molecules for each of the four sizes of Au$_m$ clusters that we have studied theoretically. Thus in an assumed top-down decay process, the systems with lower numbers $n$ of imidazole molecules in each of the panels of Figure \ref{fig:AuImMS} will likely have been depleted by the loss of Au atoms, contributing among other pathways to the strong signals from AuIm$_2^+$ and Au$_3$Im$_3^+$ that we detect.

An interesting feature that stands out in the experiments is the pair of Au$_4$Im$_3^+$ and Au$_4$Im$_4^+$, which are about equally abundant. It is not clear from the dissociation energies (Figure \ref{fig:AuIm_DFT_BEs}) why this is the case, although some clue might be given by the structures in Figure \ref{fig:Au4Imn}. We find no stable rhombic structure for Au$_4$Im$_3^+$, although there is one for Au$_4$Im$_4^+$. There could be some reaction barrier that stabilizes the rhombic Au$_4$Im$_4^+$ $\rightarrow$ Au$_4$Im$_3^+$ + Im dissociation pathway that gives the double peak, although this remains speculative.

The structures of Au$_m$Im$_n^+$ that we have identified with out calculations are distinctively different than the sandwich and riceball structures found in studies of complexes consisting of transition metals and aromatic molecules like benzene\cite{Schroder:2000aa,Nakajima:2000aa,Koyanagi:2003aa,Coriani:2006aa,Yao:2015aa,Flores:2016aa} and cyclopentadienyl.\cite{Werner:2012aa} This distinction is perhaps most clear when comparing the dumbbell AuIm$_2^+$ (structure {\bf (e)}, Figure \ref{fig:Au1Imn}) with, for example, ferrocene Fe(C$_5$H$_5$)$_2$. Ferrocene is made up of an iron atom sandwiched between two cyclopentadienyl rings with Fe forming equally distributed $\eta^5$ bonds with the carbon atoms in each molecule.\cite{Werner:2012aa} Imidazole is essentially a cyclopentadiene molecule with two nonadjacent CH groups replaced with N and NH (N$^3$ and N$^1$ in Figure \ref{fig:Im0}), respectively, and it is the lone electron pair from the N$^3$ site that gives the strongest bonds with gold. This preference is observed for all of the larger Au$_m$Im$_n^+$ systems that we have studied here and is the reason why sandwiched structures are disfavored.

It is thus clear that gold and imidazole form strong chemical bonds, so strong that imidazole is able to break down smaller gold clusters. However, a complete theoretical picture of properties of these systems will require more advanced calculations than presented here, such as dynamical simulations, to properly include reaction barriers and the role of different cluster geometries, especially for larger systems. Nonetheless, our experimental results clearly show that different sizes of gold clusters preferentially bind specific numbers of imidazole molecules. The structures of these systems are given by our calculations, which also gives insight into the stabilities of these systems and the reasons for their abundance. It will be interesting to compare these results with others where imidazole is replaced by other molecular systems, insights that will improve our understanding of the chemical nature of gold nanoparticles.

\section*{Acknowledgements}
This work was supported by the Austrian Science Fund FWF (projects P26635 and M1908-N36) and the Swedish Research Council (Contract No.\ 2016-06625).

\section*{Conflict of interest}
There are no conflicts to declare.


\balance


\bibliography{Library.bib} 
\bibliographystyle{rsc} 

\end{document}